\documentclass[%
    superscriptaddress,
    twocolumn,
    nofootinbib,
    amsmath,
    amssymb,
    aps,
    pra,
    10pt
]{revtex4-2}
\usepackage[colorlinks=true,linkcolor=teal,citecolor=Maroon,urlcolor=Maroon]{hyperref}

\usepackage[utf8]{inputenc}
\usepackage[T1]{fontenc}
\usepackage{lmodern}

\usepackage[x11names, svgnames, rgb, table, dvipsnames]{xcolor} 

\usepackage{graphicx}
\usepackage{booktabs}
\usepackage{multirow}
\usepackage{array}
\usepackage{tabularx}
\usepackage{environ}

\usepackage[caption=false,font=footnotesize]{subfig}

\usepackage{bbm} 
\usepackage{tom} %
\usetikzlibrary{calc, arrows.meta, trees, quotes, matrix, positioning, shapes, fit}
\useyquantlanguage{groups}
\newcommand{\wt}{\mathrm{wt}}

\newcommand{\condref}[1]{\hyperref[#1]{Condition~\ref*{#1}}}

\makeatletter
\DeclareRobustCommand\rvdots{%
\vbox{%
\baselineskip4\p@\lineskiplimit\z@%
\kern-\p@%
\hbox{.}\hbox{.}\hbox{.}%
}%
}
\makeatother

\SetKwComment{Comment}{/* }{ */}

\makeatletter
\newcounter{problem}[section]
\renewcommand{\theproblem}{\arabic{problem}}
\newcommand{\problemtitle}[1]{\gdef\@problemtitle{#1}}
\newcommand{\probleminput}[1]{\gdef\@probleminput{#1}}
\newcommand{\problemquestion}[1]{\gdef\@problemquestion{#1}}
\NewEnviron{problem}{
  \refstepcounter{problem}
  \problemtitle{}\probleminput{}\problemquestion{}
  \BODY
  \par\addvspace{.5\baselineskip}
  \noindent \textbf{Problem \theproblem: \@problemtitle}%
  \par\addvspace{.5\baselineskip}
  \noindent\begin{tabularx}{\linewidth}{@{\hspace{\parindent}} l X c} %
    \textbf{Input:} & \@probleminput \\
    \textbf{Problem:} & \@problemquestion
  \end{tabularx}
  \par\addvspace{.5\baselineskip}
}
\makeatother
\crefname{problem}{Problem}{Problems}

\begin{document}
\title{Synthesis of Fault-tolerant State Preparation Circuits\\using Steane-type Error Detection}

\author{Erik Weilandt}
\email{erik.weilandt@tum.de}
\affiliation{Chair for Design Automation, Technical University of Munich, Germany}

\author{Tom Peham}
\affiliation{Chair for Design Automation, Technical University of Munich, Germany}

\author{Robert Wille}
\affiliation{Chair for Design Automation, Technical University of Munich, Germany}
\affiliation{MQSC, Garching near Munich, Germany}

\begin{abstract}
    Fault-tolerant state preparation is essential for reliable quantum error correction, particularly in Steane-type error correction, which relies on robust ancilla states for syndrome readout. 
    One method of fault-tolerant state preparation is to initialize multiple ancilla states and check them against each other to detect problematic errors.
    In the worst case, the number of states required for successful initialization grows polynomially with the code distance, but it has been shown that this can be reduced to constant ancilla overhead---in the best case, only four states are required.
    However, existing techniques for finding low-overhead initialization schemes are limited to codes with large symmetry groups, such as the Golay code.
    In this work, we propose a general, automated synthesis methodology for Steane-type fault-tolerant state preparation circuits that applies to arbitrary \emph{Calderbank-Shor-Steane} (CSS) codes and does not rely on code symmetries.
    We apply the proposed methods to various CSS codes up to a distance of seven and simulate the successful fault-tolerant initialization of logical basis states under circuit-level depolarizing noise.
    The circuits synthesized using the proposed methodology provide an important step towards experimental realizations of high-fidelity ancilla states for \mbox{near-term} demonstration of fault-tolerant quantum computation.
\end{abstract}

\maketitle

\section{Introduction}
\label{sec:intro}

A major challenge on the path towards utility-scale quantum computing is the noise inherent in any quantum system.
\emph{\mbox{Fault-tolerant} quantum computing} (FTQC) and \emph{quantum error correction} (QEC) address this issue by encoding quantum information redundantly in a quantum error correction code\mbox{~\cite{kitaevFaulttolerantQuantumComputation2003,nielsenQuantumComputationQuantum2010,gottesmanTheoryFaulttolerantQuantum1998}}. 
\emph{Calderbank-Shor-Steane} (CSS) codes~\cite{shorFaulttolerantQuantumComputation1996,calderbankGoodQuantumErrorcorrecting1996,steaneErrorCorrectingCodes1996} are a prominent class of stabilizer codes~\cite{gottesmanStabilizerCodesQuantum1997} that are constructed from two classical linear codes which allow for independently detecting and correcting bit-flip (Pauli $X$ errors) and phase-flip (Pauli $Z$ errors).

An essential property of CSS codes is that any CSS code admits a transversal CNOT gate, which means that a logical CNOT gate between the qubits of two code blocks can be implemented via physical CNOT gates between the physical qubits of the respective blocks.
Transversal gates are desirable for FTQC because errors in those gates do not spread between data qubits of the same block.
In \emph{Steane-type error correction}~\cite{steaneActiveStabilizationQuantum1997,postlerDemonstrationFaultTolerantSteane2024}, transversal CNOT gates are used to copy errors from an encoded data block to an ancillary system that encodes the logical all-zero $\ket{0}_L^{\otimes k}$ or logical all-plus $\ket{+}_L^{\otimes k}$ state. 
Measurements of the ancilla system are then used to infer and correct errors on the data qubits.
Since the ancilla system can be initialized independently and has only minimal interaction with the data block, Steane-type QEC is a promising candidate for practical quantum error correction, especially for quantum computing architectures that exhibit a high degree of gate-level parallelism like trapped-ion quantum computers~\cite{mosesRaceTrackTrappedIonQuantum2023,pinoDemonstrationTrappedionQuantum2021,figgattParallelEntanglingOperations2019} or neutral atom quantum computers~\cite{saffmanQuantumComputingAtomic2016, morgadoQuantumSimulationComputing2021, levineParallelImplementationHighFidelity2019,bluvsteinLogicalQuantumProcessor2024}.

\begin{figure*}[t]
\centering

\subfloat[\label{fig:steane-ftsp-naive}]{%
    \includegraphics[width=0.69\linewidth]{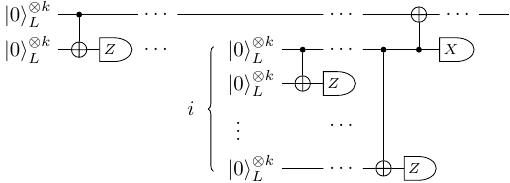}%
}
\hfill %
\subfloat[\label{fig:steane-ftsp-opt}]{%
    \includegraphics[width=0.30\linewidth]{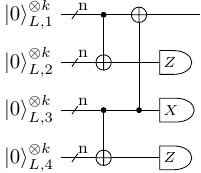}%
}

\caption{Non-deterministic Steane-type state preparation. \textbf{(a)} The state is prepared by non-fault-tolerantly preparing multiple states and copying errors on data qubits over to ancilla qubits, which are measured out in the respective basis for error detection. This process is repeated until no errors are detected on any of the states. This process requires $O(nd^2)$ physical qubits and measurements if the states are prepared using the same circuit, since the ancilla needs to be recursively verified using additional ancillas. \textbf{(b)} Four states are prepared using different state preparation circuits. This construction is fault-tolerant if the sets of propagated errors of all circuits are sufficiently different.}
\label{fig:steane-type-prep}

\end{figure*}

The ancillary systems used for Steane-type QEC should have a high fidelity to avoid introducing further errors on the data. 
This necessitates robust, fault-tolerant state preparation circuits.
One way to create high-fidelity states is via a non-deterministic \emph{repeat-until-success} protocol, where the state is initialized with a non-fault-tolerant unitary encoding circuit augmented by an error-detection gadget. 
The initialization is then repeated until the error-detection circuit measurements indicate that no problematic errors occurred during state preparation.

One way to implement the error-detection circuit is to measure specific stabilizers of the code~\cite{gotoMinimizingResourceOverheads2016,pehamAutomatedSynthesisFaultTolerant2025,bluvsteinLogicalQuantumProcessor2024,reichardtLogicalComputationDemonstrated2024}.
While manageable for small codes, the number of measurements required to detect all problematic errors quickly becomes large. Hence, it is increasingly more difficult to find such a minimal set of measurements~\cite{pehamAutomatedSynthesisFaultTolerant2025}.
Furthermore, \mbox{high-weight} stabilizer measurements need to be protected by flag qubits~\cite{chamberlandFlagFaulttolerantError2018,chaoFlagFaultTolerantError2020} to avoid introducing hook errors on the data qubits.
Yet another method is to incorporate flag qubits directly into the unitary encoding circuit~\cite{forlivesiFlagOriginModular2025}. This approach reduces circuit overhead for error detection, since errors are detected within specific subcircuits during encoding rather than at the end of the circuit. However, this method requires a specific circuit structure with significant depth overhead.

An alternative state preparation approach is illustrated in ~\Cref{fig:steane-ftsp-naive}. One prepares multiple ancilla states, copies potential errors from one ancilla state to another using a transversal CNOT gate, and detects them by measuring the second state. 
As with Steane-type QEC, this has the benefit of a low-depth overhead since the ancilla can be prepared in parallel.
The issue is that specific errors might cancel out when they are copied with the transversal CNOT gate, effectively lowering the distance of the code. 
Consequently, multiple verifications are required to ensure no such errors are undetected.

Ref.~\cite{paetznickFaulttolerantAncillaPreparation2013} shows how this ancilla overhead can be made constant for the Golay code by preparing four states using different preparation circuits that guarantee that high-weight errors do not go by undetected due to cancellations on the logical ancilla qubits as illustrated in~\Cref{fig:steane-ftsp-opt}.
This approach is based on qubit permutations using qubit symmetries but has been shown only for the highly symmetrical Golay code~\cite{paetznickFaulttolerantAncillaPreparation2013}, Reed-Muller codes~\cite{gongComputationQuantumReedMuller2024}, and GHZ ancilla states~\cite{Peham_2026} thus far; no general circuit-synthesis methodology has been proposed for constructing constant-overhead Steane-type fault-tolerant state preparation circuits (FTSPs).

This work closes this gap by proposing a general circuit synthesis methodology for this state preparation scheme. 
We show that constructing the desired circuits does not rely on code symmetries.
In fact, the problem can be addressed by using different CNOT circuits that prepare the same state but propagate errors differently.
Code symmetries are then only one specific way of obtaining such circuits.

Our main contributions are as follows:

\begin{itemize}
    \item We explicitly give manual constructions for fault-tolerant \mbox{Steane-type} state preparation circuits for the $\llbracket19,1,5\rrbracket$ code, which has only a few qubit symmetries, and the $\llbracket17,1,5\rrbracket$ code, which has no qubit symmetries at all.
    \item For circuits with combinatorially more challenging error sets, we utilize a heuristic CNOT circuit synthesis that constructs multiple circuits with different fault sets on the fly.
    \item Using the proposed automated synthesis method, we construct several low-depth Steane-type fault-tolerant state preparation schemes for codes up to distance 7.
\end{itemize}

The proposed methodology is intended for the regime of small- and
medium-distance codes where explicit enumeration of fault sets remains
tractable. This regime aligns with current and near-term experimental
demonstrations of fault-tolerant quantum computing, where codes of
distances up to seven are being explored on trapped-ion and neutral-atom
platforms. For larger distances, the size of the fault sets to be checked
during synthesis grows quickly and the method becomes increasingly
difficult to apply.

We compare the resulting circuit designs with state preparation circuits from Ref.~\cite{pehamAutomatedSynthesisFaultTolerant2025} as well as Ref.~\cite{forlivesiFlagOriginModular2025}, and provide numerical evidence for the fault tolerance properties of the synthesized circuits
via Monte Carlo simulation.
All proposed methods and circuits are publicly available in the form of open-source software as part of the \emph{Munich Quantum Toolkit} (MQT~\cite{willeMQTHandbookSummary2024}).

The rest of this manuscript is structured as follows. \Cref{sec:background} provides the necessary background on FTQC and introduces the noise model and fault tolerance criteria considered in this work. \Cref{sec:motivation} reviews and motivates the problem of circuit synthesis for Steane-type fault-tolerant state preparation, followed by~\Cref{sec:states}, which shows explicit constructions for distance five color codes.
\Cref{sec:heuristic} proposes an automated synthesis method for constructing Steane-type fault-tolerant state preparation circuits.
\Cref{sec:eval} shows the results of conducted simulations, which show that the proposed constant low-overhead constructions are strictly fault-tolerant.
Finally, \Cref{sec:conclusion} concludes this paper.

\section{Background}
\label{sec:background}

This section provides background on the relevant concepts of quantum error correction and establishes the notation used in this work.

Throughout this work we use $[i] = \{1,2,\dots, i\}\subseteq \Z$.

\subsection{CSS Codes}
\label{sec:quantum-codes}
 
An $\llbracket n,k,d\rrbracket$ stabilizer code on $n$ qubits is defined by a stabilizer group $\calS$---a commutative subgroup of the $n$-qubit Pauli group $\calP_n$---which encodes $k$ logical qubits.
The code space of a stabilizer code is defined as the simultaneous $+1$ eigenspace of all elements of the stabilizer group. The distance of the code is defined as the minimal-weight Pauli operator that acts non-trivially on the code space, where we define the weight $\wt(P)$ of a Pauli operator $P$ on $n$ qubits as the number of qubits on which it acts non-trivially.
We say that two \mbox{Pauli-operators} $P_1,P_2$ are \emph{stabilizer-equivalent} (or equivalent \emph{up to stabilizers}) with respect to $\calS$, denoted $P_1\simeq_\calS P_2$, if there is an element $S\in \calS$ such that $S P_1=P_2$.

A stabilizer code is CSS if it is the combination of two groups $\calS_X, \calS_Z$, where all elements of $\calS_X$ ($\calS_Z$) are tensor products of Pauli $X$ ($Z$) and the identity $I$.
The groups $\calS_X$ and $\calS_Z$ can be identified with $\F_2$ vector spaces and are often described via the \emph{check matrices} $H_X$ and $H_Z$ whose rows are a basis of the respective vector spaces.
The CSS condition states that $H_XH_Z^T=0$, which expresses the notion that all elements of $\calS_X$ and $\calS_Z$ commute.

A detectable error (or just an error) is a Pauli operator that moves a code word out of the code space.
For CSS codes, we usually treat $X$-type and $Z$-type errors separately, even if the noise includes combinations of both. This is because CSS codes can detect and correct $X$- and \mbox{$Z$-type} errors independently, and any mixed error can be broken down into a combination of these two types and the identity $I$.
An $X$-error $E\in \F_2^n$ is detected by the $Z$-checks, i.e., $H_ZE \neq 0$ (similar for $Z$-errors and $X$-checks).
Logical $X$ and $Z$ operators $L_X, L_Z$ are defined as operators that commute with all stabilizers and anticommute with each other.

\subsection{Circuit-Level Noise Model}
\label{sec:circuit-level-noise}

This work considers quantum circuits composed of physical qubit initializations in $\ket{0}$ and $\ket{+}$, CNOT gates between arbitrary qubits, and measurements in the $X$ or $Z$ basis.
We employ a circuit-level noise model parameterized by a \emph{physical error rate} $p$.
Noisy gates are modeled as ideal gates followed by a depolarizing noise channel parameterized by a physical noise rate $p$:
\begin{align*}
  \epsilon_2(\rho) &= (1-p)\rho + \frac{p}{15} \sum_{E \in \mathcal{E} }E \rho E.
\end{align*}

Here, $\mathcal{E} = \{P_1\otimes P_2 \mid P_1,P_2 \in \{I,X,Y,Z\} \}\setminus \{I\otimes I\}$. 
In addition, qubits are initialized in the $-1$ eigenstate of the respective basis with probability $O(p)$, and measurements also flip with a probability of $O(p)$.
If a qubit is idle during one layer of CNOT gates, it is also subject to single-qubit depolarizing \enquote{idling} noise of strength $O(p)$.

The notion of the types of errors that can propagate through a circuit via two-qubit gates is central to the following discussions:

\begin{definition}\label{def:fault-set}
    Let $\calC$ be a quantum circuit with $n$ qubits and $m$ gates, and let $\calC_j$ denote the circuit obtained from $\calC$ by removing the first $j$ gates. 
    The $X$ ($Z$) \emph{fault set} $\calE_X(\calC)$ ($\calE_Z(\calC)$) of $\calC$ is the set of errors stemming from a single propagated $X$ ($Z$) fault in $\calC$. 
    More precisely \mbox{$\calE_X(\calC) = \{\calC_j X_i \calC_j^{-1} \mid i\in [n], j \in [m]\}$}. 
\end{definition}

The \emph{minimal weight} of an error $E$ is defined as $\min_{S\in\calS}\wt(S\cdot E)$.
We will generally omit the stabilizer group $\calS$ and, since we are only concerned with minimal representatives, refer to the minimal weight simply as $\wt(E)$.

As we aim to eventually construct fault-tolerant state preparation circuits, we define strict fault tolerance: a circuit should not amplify errors during execution.

\begin{definition}\label{def:fault-tolerance}
  We say that a state preparation circuit $\calC$ for an $\llbracket n,k,d\rrbracket$ code is \emph{strictly fault-tolerant} if there are no errors $e_1, \cdots, e_t \in \calE_X(\calC)$ ($\calE_Z(\calC)$), $t \leq \floor{(d-1)/2}$ such that $\wt(\sum_{i=1}^te_i) > t$.
\end{definition}

\section{Problem Definition and Motivation}
\label{sec:motivation}

We consider the problem of initializing the logical $\ket{0}_L^{\otimes k}$ state of an $\llbracket n,k,d\rrbracket$ CSS code.
One way to achieve this is by initializing the $n$ physical qubits to $\ket{0}$ and $\ket{+}$ and executing a unitary CNOT encoding circuit to entangle these qubits into the desired state.
Such an encoding circuit is generally not \mbox{fault-tolerant}, as single Pauli errors in the circuit can propagate through a CNOT gate:

\begin{center}
\includegraphics[width=.8\linewidth]{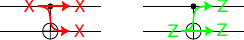}
\end{center}

One approach to making unitary state preparation resilient to such errors is by initializing two copies of a state, using one state to verify whether errors have occurred in the other state. 
In the case of CSS codes, this can be done with a simple circuit using transversal CNOT gates and measurement of two ancilla blocks, one measured in the Z basis to detect X-type errors, one in the X basis to detect Z-type errors:

\begin{center}
\includegraphics[width=.7\linewidth]{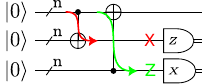}
\end{center}

Since a transversal CNOT gate on a CSS code implements a logical CNOT gate between all encoded qubits of two code blocks, this circuit acts trivially on the code space. 
However, errors still propagate through the physical CNOT gates. 
These errors can be detected in the ancilla measurements. 
Therefore, this circuit can be used in a \mbox{\emph{repeat-until-success}} scheme by post-selecting on these measurements.
We refer to such a state preparation scheme as \emph{Steane-type (non-deterministic) fault-tolerant state preparation} (FTSP).

\begin{figure}[t]
    \centering
    \includegraphics[width=\linewidth]{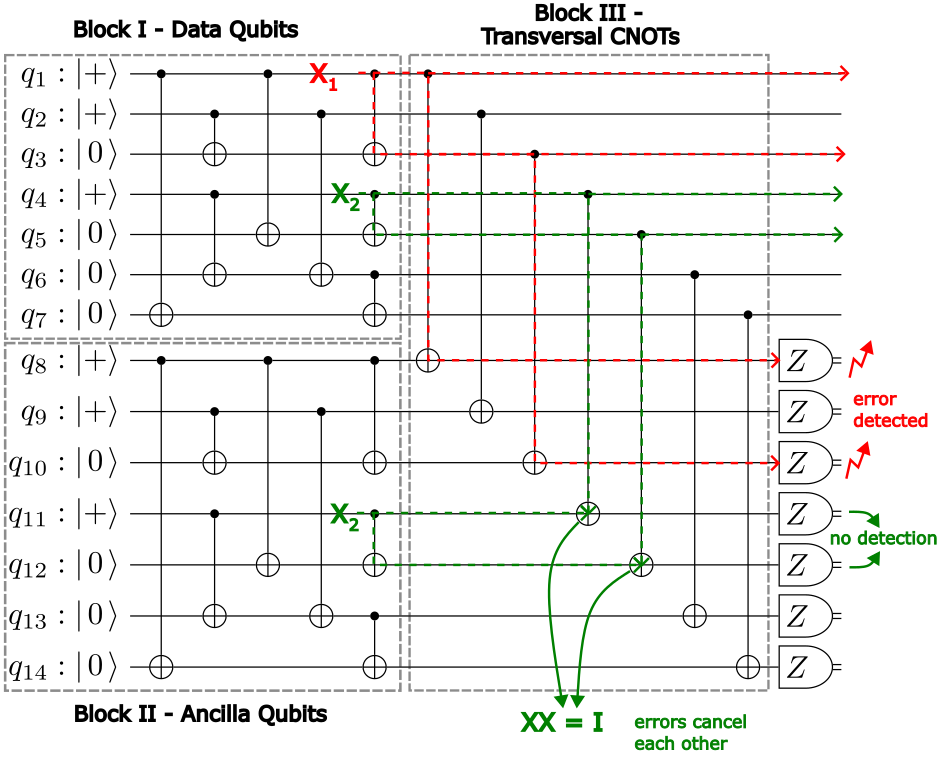}
    \caption{Block I and II are state preparation circuits for the logical zero state of the code. Block I represents the data structure, and Block II is the ancilla structure. Block III is the connection of both state preparation circuits through transversal CNOT gates. Together, all three blocks form the base structure of Steane-type error correction.}
    \label{fig:steane-code-example}
    \vspace*{-3mm}
\end{figure}

\begin{example}
    Consider Steane-type preparation of the logical state $\ket{0}_L$ for the Steane code, as shown in \Cref{fig:steane-code-example}. In this scenario, the data block (Block I) undergoes a single-qubit error $X_1$, which propagates within Block I to a weight-2 error. During the subsequent transversal CNOT operation (Block III), this error is further transferred from the data block to the ancilla block (Block II). As a result, the error remains on data qubits $q_1$ and $q_3$ and on the corresponding ancilla qubits $q_8$ and $q_{10}$, where it is ultimately detected through measurement of the ancilla.
\end{example}

Depending on how the state preparation circuits are chosen in Steane-type FTSP, the protocol may fail to be fault-tolerant. The issue arises when the same preparation circuit is used for both the data and the ancilla blocks: identical faults occurring at the same circuit location can propagate through the transversal CNOT in such a way that they cancel on the ancilla. In this situation, the ancilla measurements reveal no syndrome, even though a nontrivial error remains on the data block. Thus, the data error goes undetected, breaking fault tolerance.

\begin{example}
Consider the circuit in \Cref{fig:steane-code-example} again. If an $X_{2}$ error occurs at the same location in both state preparation circuits, it propagates through the transversal CNOT from the data block to the ancilla block, where it cancels with the identical error already present on the ancilla. As a result, the ancilla measurements reveal no syndrome, leaving the error on the data block undetected.

While this cancellation does not affect the Steane code---which only protects against single-qubit errors---it becomes problematic for larger, higher-distance codes, where such undetected cancellations violate strict fault tolerance.
\end{example}

A brute-force way to compensate for such errors is to use additional ancillas and repeat the transversal CNOT gate and measurement multiple times.
Unfortunately, the ancillas used to verify the data can introduce high-weight errors, so they would need to be checked by additional ancillary qubits. 
In fact, ancillas need to be recursively verified until it is ensured that all occurrences of $t < \frac{d-1}{2}$ independent errors can be detected.
This is the protocol depicted in~\Cref{fig:steane-ftsp-naive}.

With increasing distance, this protocol quickly results in circuits with high depth and CNOT gate count since many states are required for the recursive verification scheme.
Because each additional ancilla block adds CNOT gates and qubits, each of which is itself a source of new faults, the recursive verification scheme has a large number of fault locations. Moreover, each level of recursion further reduces the acceptance rate.
This extremely pessimistic scheme ensures that, even if errors cancel out on an ancilla, another check will still detect the error.

Ref.~\cite{paetznickFaulttolerantAncillaPreparation2013} shows that four state preparation circuits are sufficient for Steane-type FTSP, provided the circuits are chosen so that their fault sets are sufficiently distinct to prevent the problematic error cancellations described above (see~\Cref{fig:steane-ftsp-opt}).
The notion of sufficiently distinct fault sets is formalized below: no nontrivial error can cancel during any transversal CNOT.
In this construction, the measurement of the second ancilla $\ket{0}_{L,2}^{\otimes k}$ detects all $X$-type faults that have propagated from the first state $\ket{0}_{L,1}^{\otimes k}$.
Similarly, the third ancilla $\ket{0}_{L,3}^{\otimes k}$ detects all propagated $Z$-type faults.
Finally, the fourth ancilla $\ket{0}_{L,4}^{\otimes k}$ detects any $X$ faults on $\ket{0}_{L,3}^{\otimes k}$, ensuring that the $Z$-syndrome extraction on $\ket{0}_{L,1}^{\otimes k}$ does not introduce new undetected errors.
If all three verification steps succeed, the remaining state $\ket{0}_{L,1}^{\otimes k}$ has residual error weight at most $(d-1)/2$, provided that no more than $(d-1)/2$ faults occurred during preparation.
This guarantee is achieved while requiring only a \emph{constant} number of ancillas.

Achieving this constant overhead requires choosing circuits whose fault sets are sufficiently distinct. %
To formalize what it means for two circuits to be “different enough,” we introduce the notion of \emph{$t$-distinctness}.
Intuitively, two fault sets are $t$-distinct if no combination of at most $t$ faults from each set can produce equivalent errors on both circuits, preventing cancellations under transversal interaction, and if any such combination necessarily has weight larger than the number of faults involved.

\begin{definition}\label{def:t-distinct}
Let $\calE_1, \calE_2$ be two fault sets and $\calS$ be a stabilizer group. 
Then, we say that $\calE_1, \calE_2$ are \emph{$t$-distinct} with respect to $\calS$ if there are no subsets $\calF_1 \subseteq \calE_1, \calF_2 \subseteq \calE_2$ with $|\calF_1| + |\calF_2| \leq t$ such that \mbox{$\prod_{e\in \calF_1}e \simeq_\calS \prod_{e\in \calF_2}e$} and \mbox{$\wt\prod_{e\in \calF_1}e>|\calF_1| + |\calF_2|$}.
\end{definition}

At first glance, one might expect that all fault sets of the circuits $\mathcal{C}_1, \mathcal{C}_2, \mathcal{C}_3, \mathcal{C}_4$ preparing $\ket{0}_{L,1}$, $\ket{0}_{L,2}$, $\ket{0}_{L,3}$, and $\ket{0}_{L,4}$ must be $\lfloor (d-1)/2 \rfloor$-distinct in order for the scheme in~\Cref{fig:steane-ftsp-opt} to be strictly fault-tolerant. However, this is not necessary. Examining how errors propagate and interact in~\Cref{fig:steane-ftsp-opt} shows that only certain pairs of fault sets must satisfy $\lfloor (d-1)/2 \rfloor$-distinctness.

We say that four state preparation circuits $\mathcal{C}_1,\mathcal{C}_2,\mathcal{C}_3,\mathcal{C}_4$ for an $\llbracket n,k,d\rrbracket$ CSS code implement the protocol in~\Cref{fig:steane-ftsp-opt} in a strictly fault-tolerant manner if the following pairs of fault sets are $\lfloor(d-1)/2\rfloor$-distinct:

\begin{align}
  1)\quad & \mathcal{E}_X(\mathcal{C}_1)                      && \mathcal{E}_X(\mathcal{C}_2) \nonumber \\
  2)\quad & \mathcal{E}_X(\mathcal{C}_3)                      && \mathcal{E}_X(\mathcal{C}_4) \label{eq:steane-ftsp-condition} \\
  3)\quad & \mathcal{E}_Z(\mathcal{C}_1)\cup \mathcal{E}_Z(\mathcal{C}_2)  && \mathcal{E}_Z(\mathcal{C}_3)\cup \mathcal{E}_Z(\mathcal{C}_4) \nonumber
\end{align}

This work aims to construct such circuits for arbitrary CSS codes, as summarized by the following problem statement.

\begin{problem}\label{prob:steane-ftsp}
    \problemtitle{Circuit Synthesis for Steane-type FTSP}
    \probleminput{Stabilizer generators of an $\llbracket n,k,d\rrbracket$ CSS code}
    \problemquestion{Construct four state preparation circuits $\calC_1, \calC_2, \calC_3, \calC_4$ for $\ket{0}_L$ that implement the non-deterministic FTSP shown in~\Cref{fig:steane-ftsp-opt}.}
\end{problem}

The $t$-distinctness requirement necessitates leveraging the degrees of freedom in circuit construction to alter the circuits' fault sets.
In Ref.~\cite{paetznickFaulttolerantAncillaPreparation2013}, four $3$-distinct state preparation circuits are constructed for the Golay code~\cite{steaneSimpleQuantumErrorcorrecting1996} by exploiting the Golay code's large symmetry group, the Mathieu group $M_{23}$.
The circuits are obtained by taking a circuit $\calC_1$ and deriving the three other circuits by applying certain symmetries of the code.
Such permutations preserve the encoded logical information while altering the specific fault set associated with a state preparation. 

A key limitation of this approach is that it relies on large symmetry groups to achieve $t$-distinctness. 
While powerful, this technique does not apply to all CSS codes, as many relevant codes lack sufficiently large symmetry groups. 
Consequently, a more generalized approach is required---one that does not rely on code symmetries but achieves $t$-distinctness by appropriately constructing the CNOT circuits. 
In this work, we will show how this can be achieved in an automated fashion for codes of moderate size that are suitable for Steane-type FTSP.

\begin{figure*}[h!t]
    \centering
    
    \subfloat[$\llbracket19,1,5\rrbracket$ color code\label{fig:cc-666}]{%
        \includegraphics[width=0.4\linewidth]{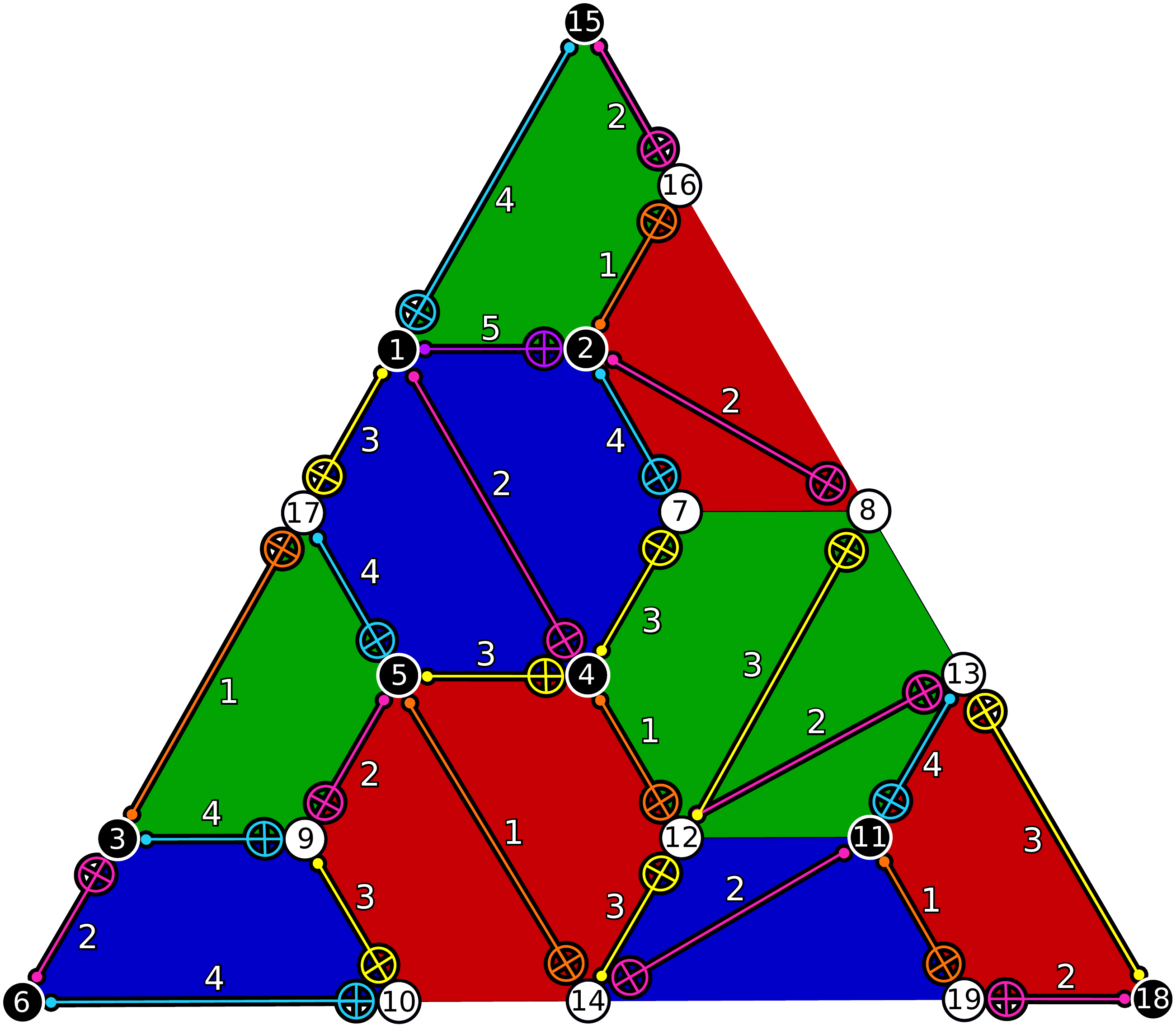}%
    }
    \hfill
    \subfloat[$\llbracket17,1,5\rrbracket$ color code\label{fig:cc-488}]{%
        \includegraphics[width=0.55\linewidth]{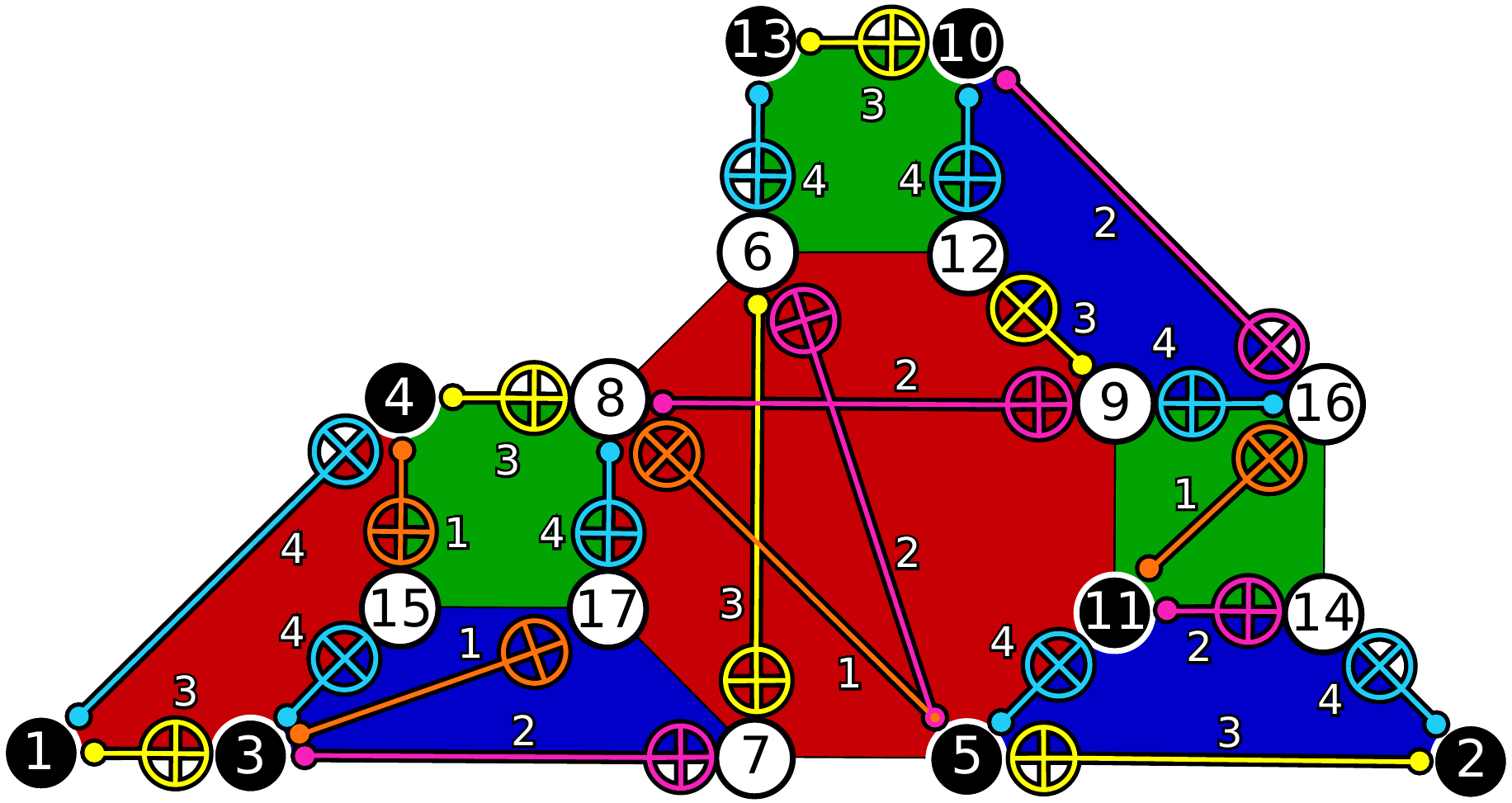}%
    }
    
    \caption{
        State preparation circuits for the $\ket{0}_L$ of the $\llbracket19,1,5\rrbracket$ and $\llbracket17,1,5\rrbracket$ color codes~\cite{bombinTopologicalQuantumDistillation2006}.
        Black vertices represent qubits prepared in $\ket{+}$, and white vertices represent qubits prepared in $\ket{0}$.
        The CNOT order is indicated by the edge colors (orange, magenta, yellow, blue, and purple).
    }
    \label{fig:color-code-prep}
    \vspace*{-4mm}
\end{figure*}

\section{Manual Constructions for Steane-Type Fault-Tolerant State Preparation}
\label{sec:states}

It is impractical to independently synthesize the circuits $\calC_1, \calC_2, \calC_3, \calC_4$ in order to solve~\Cref{prob:steane-ftsp}, because the $t$-distinctness constraints tightly couple their fault sets.
Intuitively, a much better approach is to take a \emph{reference circuit}, say $\calC_1$, and generate the other circuits by altering this reference circuit.

The following two examples show how symmetries or local circuit rewrites can be used to obtain circuits with suitably distinct fault sets.
From this, we derive circuit rewriting techniques that allow us to synthesize Steane-type FTSP circuits for higher-distance codes that are hard to reason about manually.

\subsection{$\llbracket19,1,5\rrbracket$ Color Code}
\label{sec:19-1-5}

The $\llbracket19,1,5\rrbracket$ color code shown in~\Cref{fig:cc-666} is obtained from a hexagonal tiling of the plane. 
Qubits are placed on the vertices of a triangular region of the tiling, and the red, green, and blue plaquettes represent both $X$ and $Z$ stabilizers.
This code's symmetry group is generated by a single element: a clockwise rotation by $120$ degrees:
\[\setlength\arraycolsep{1.2pt}
  \begin{pmatrix}
  1 & 2 & 3 & 4  & 5  & 6 & 7 & 8  & 9 & 10 & 11 & 12 & 13 & 14 & 15 & 16& 17 & 18 & 19\\
  10 & 9 & 19 & 4 & 12 & 18 & 5 & 17 & 11 & 13  & 2 & 7 & 1  & 8 & 6 & 3 & 14 & 15 &16 
\end{pmatrix}.
\] 
These symmetries allow us to generate additional state-preparation circuits by permuting qubits while preserving the logical state.

A state preparation circuit $\calC_1$ for the $\ket{0}_L$ with a CNOT depth of $5$ and $27$ CNOT gates is shown by the CNOT circuit in~\Cref{fig:cc-666}. 
Note that no CNOT gates act on qubits that do not have support on the same stabilizer generator. 
It follows immediately that CNOT gates on the qubits of the weight-four stabilizers cannot propagate an error of weight greater than $2$.

Furthermore, the CNOT gates on the weight-six plaquettes only propagate errors of at most weight $2$, namely 
\[\{X_1X_2, X_4X_5, X_{12}X_{14}, X_9X_{10}, X_4X_7, X_8X_{12}, X_{11}X_{13}\}.\]
Therefore, $\calE_X(\calC_1)$ is $2$-distinct to itself. 
Setting 
\[\mathcal{C}=\mathcal{C}_1 = \mathcal{C}_2 = \mathcal{C}_3 = \mathcal{C}_4\]
would therefore already fulfill requirements 1) and 2) of~\condref{eq:steane-ftsp-condition}.

To fully achieve fault tolerance, we must also ensure \mbox{2-distinctness} for Z errors between the circuit pairs $\{\mathcal{C}_1, \mathcal{C}_2\}$ and $\{\mathcal{C}_3, \mathcal{C}_4\}$. 
There is (up to stabilizer) only one propagated $Z$ error of weight $3$, namely $Z_4Z_5Z_{17}$.
Applying a counter-clockwise rotation $\sigma$ acting on the qubits of $\calC_1$ maps this error to $Z_4Z_{12}Z_{14}$, while applying a clockwise rotation $\sigma^2$ maps it to $Z_4Z_{12}Z_{11}$.
Setting 
\[\calC_2 = \calC_1, \calC_3 = \calC_4 = \sigma(\calC_1)\]
therefore yields four circuits that fulfill all the prerequisites of ~\condref{eq:steane-ftsp-condition}.
\subsection{$\llbracket17,1,5\rrbracket$ Color Code}
\label{sec:17-1-5}

Unlike the $\llbracket19,1,5\rrbracket$ code, the $\llbracket17,1,5\rrbracket$ code shown in~\Cref{fig:cc-488} has only a trivial qubit symmetry. Thus, we cannot rely on permutations to obtain distinct fault sets and must instead modify the CNOT structure directly. 
Nevertheless, the same key arguments as with the $\llbracket19,1,5\rrbracket$ hold: errors propagating on the weight-4 stabilizers can be ignored, and our focus shifts to the single weight-8 stabilizer.
The state preparation circuit $\calC_1$ in~\Cref{fig:cc-488} has (up to stabilizers) only a single $X$ error with weight $>2$, namely $X_5X_6X_7X_{11}$.
All $Z$ errors have a weight of at most two, so $\calE_Z(\calC_1)$ is $2$-distinct to itself.

We can construct a circuit with a different $X$ fault set by altering the CNOT circuit on the weight-eight plaquette.
One way to do this is to remove $\mathrm{CX}_{5,11}$ and add $\mathrm{CX_{9,11}}$ before $\mathrm{CX}_{9,16}$.
This local modification yields a circuit $\calC_2$ with single propagated $X$ errors with \mbox{weight $>2$} of $\{X_5X_6X_7, X_{10}X_{11}X_{16}\}$. 
This set is $2$-distinct to $\{X_5X_6X_7X_{11}\}$ so a Steane-type FTSP circuit can be constructed by setting $\calC_3=\calC_1, \calC_4=\calC_2$.

\section{Automated Synthesis of Steane-type Fault-tolerant State Preparation Circuits}
\label{sec:heuristic}

\begin{figure*}[h!tb]
  \centering
\includegraphics[width=\linewidth]{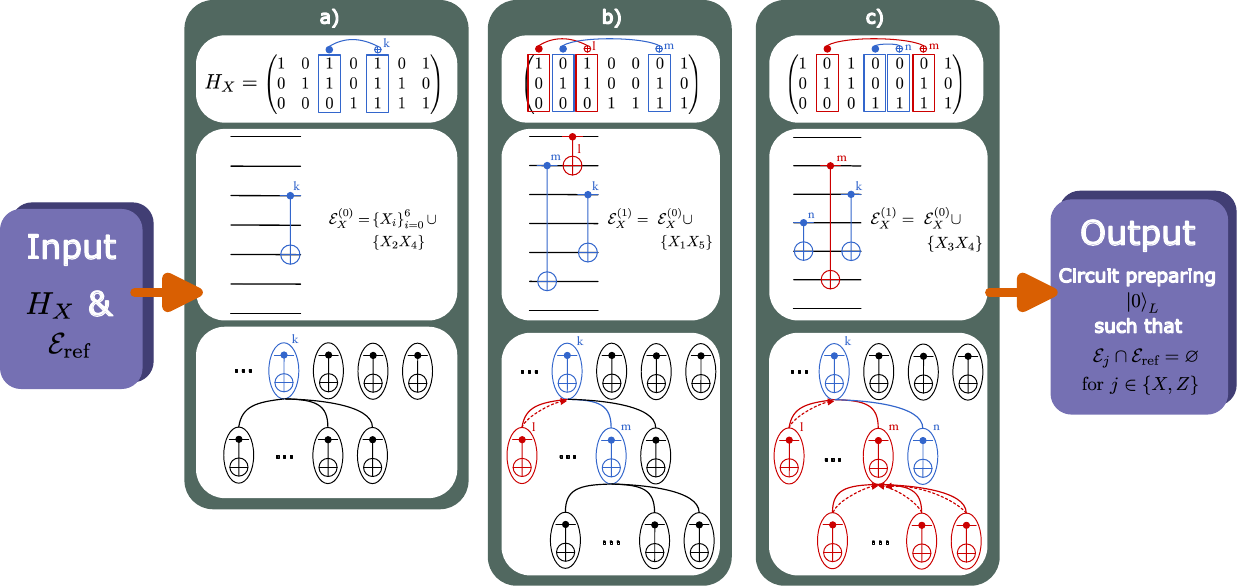}
\caption{
Fault-set–guided synthesis of a state-preparation circuit. 
Given the check matrix $H_X$ and a reference fault set $\calE_{\text{ref}}$, the algorithm constructs a circuit whose propagated fault set is $t$-distinct from $\calE_{\text{ref}}$. 
Panels (a–c) illustrate three different scenarios in the search process: the evolving check matrix (top), the partially constructed circuit assembled from right to left (middle), and the remaining candidate CNOTs (bottom). 
Blue CNOTs satisfy the $t$-distinctness constraint, while red CNOTs violate it and are therefore pruned.
\textbf{(a)} A valid CNOT is selected according to the cost function, and the fault set is updated.  
\textbf{(b)} A candidate CNOT violates $t$-distinctness and is discarded in favor of an alternative choice.  
\textbf{(c)} All candidates violate $t$-distinctness, forcing the algorithm to backtrack by removing previously placed CNOTs.
}

\label{fig:algorithm_sketch}
\end{figure*}

Manual reasoning methods, which work well for smaller codes, as seen in
\Cref{sec:states}, become increasingly difficult to apply as the code distance
grows. For larger codes, fault sets can become enormous, and manually
analyzing them quickly becomes infeasible. The same scaling affects
the automated synthesis presented in this section: the cost of checking
t-distinctness grows with the size of the involved fault sets, and the
proportion of CNOT choices that preserve t-distinctness decreases as
the constraints tighten. The method is therefore best suited to codes
of moderate distance, where the fault sets are large enough to require
automation but small enough that explicit enumeration remains feasible.

\Cref{sec:states} showed how a Steane-type FTSP circuit can be constructed by starting with a reference circuit and slightly altering it to generate another circuit with a t-distinct fault-set. 
While this worked in the two examples shown, it is not generally possible to obtain a t-distinct circuit from a reference circuit only by local transformations.
Our goal now is to automate this process by synthesizing a circuit whose fault set is $t$-distinct from a reference fault set by construction.
To guarantee t-distinctness, we need control over the fault-sets during synthesis so we can make informed choices about appropriate CNOTs directly during circuit synthesis. 

In this section, we will describe such a \emph{fault-set guided} synthesis approach. 
Before we get to that, we briefly review how to construct state preparation circuits from a code check matrix.

\subsection{Circuit Synthesis via Gaussian Elimination}
A unitary $\ket{0}_L$ state preparation circuit for an $\llbracket n,k,d\rrbracket$ code with $H_X\in \F_2^{m_X\times n}$ and $H_Z\in \F_2^{m_Z\times n}$ is constructed by initializing $m_X$ qubits in $\ket{+}$ and $m_Z$ qubits in $\ket{0}$ followed by a CNOT circuit that transforms the check matrices of the initial product state to $H_X$ and $H_Z$. 
The check matrices of the product state $H_X^0\in \F_2^{m_X\times n}$, $H_Z^0\in \F_2^{m_Z\times n}$ are matrices that are non-zero in exactly $m_X$ and $m_Z$ columns, respectively, such that the CSS condition holds.
Applying a CNOT between qubits $i$ and $j$ to a CSS state adds column $i$ to column $j$ of the state's $X$ check matrix and column $j$ to column $i$ in the state's $Z$ check matrix.

The synthesis task asks for a sequence of CNOT gates that transform $H_X$ to $H_X^0$ and $H_Z$ to $H_Z^0$. 
This can be done via Gaussian elimination on the column space of the check matrix $H_X$.
This elimination is also guaranteed to reduce $H_Z$ as well by the CSS condition~\cite{pehamAutomatedSynthesisFaultTolerant2025}.

The state preparation circuit is then obtained by initializing the qubits corresponding to non-zero columns in $H_X^0$ to $\ket{+}$ and the others to $\ket{0}$ and running the elimination sequence in reverse---constructing the check matrices of the code from the check matrices of the product state.
Since errors propagate backward through a circuit, building the circuit from the final layers first allows us to compute partial fault sets incrementally.

Naturally, the synthesized state preparation circuits should be minimal in some sense---either being low-depth or using as few CNOT gates as possible.
There is no known efficient algorithm for synthesizing the minimal circuit, so optimal synthesis approaches use techniques like satisfiability solving to find the optimal circuit~\cite{pehamAutomatedSynthesisFaultTolerant2025}.
However, efficient heuristics also exist based on greedy best-first search~\cite{pehamAutomatedSynthesisFaultTolerant2025,websterHeuristicOptimalSynthesis2025} or A* search~\cite{websterHeuristicOptimalSynthesis2025}.

\subsection{Fault-set Guided Synthesis}

We are going to base our approach on the greedy best-first search proposed in Ref.~\cite{pehamAutomatedSynthesisFaultTolerant2025}.
This search evaluates the cost of every possible CNOT at each step by counting the number of non-zero elements in the resulting check matrix after applying it. 
The algorithm then greedily chooses the CNOT that eliminates the most non-zero entries.
Multiple CNOTs might have the same costs in this search, in which case a random CNOT is picked.

These degrees of freedom can be exploited to construct $t$-distinct circuits relative to a chosen reference circuit. 
As a first step, we synthesize such a reference circuit. Since there are no constraints on its fault set, we can obtain it using the heuristic from Ref.~\cite{pehamAutomatedSynthesisFaultTolerant2025}.
This circuit then serves as the baseline against which we construct the remaining circuits using fault-set guided synthesis.

A crucial feature of the heuristic construction is that it builds the CNOT circuit in reverse---starting from the final layers and working backward.
Since faults propagate in the opposite direction, the fault set of a circuit under construction can already be partially computed \enquote{on the fly}. 
Let $\calC_k$ denote the intermediate circuit after $k$ CNOTs have been placed.
If the intermediate CNOT circuit $\calC_k$ has fault set $\calE(\calC_k)$, then the final circuit $\calC$ will have a fault set $\calE(\calC_k) \subseteq \calE(\calC)$.
In particular, adding the gate $\mathrm{CX}_{ij}$ to $\calC_k$ yields a circuit with fault sets $\calE_X(\calC_k) \cup \{\calC_k X_iX_j \calC_k^{-1}\}$ and $\calE_Z(\calC_k) \cup \{\calC_k Z_iZ_j \calC_k^{-1}\}$.

Crucially, any additional CNOT gates inserted later in the search can only contribute more faults---since the fault set is always computed with respect to the stabilizer group of the code and not of the intermediate state, no faults are ever removed from the intermediate fault set. 

This \enquote{on-the-fly} construction of the evolving fault sets can be used to guide the search towards constructing t-distinct circuits.
Given a reference fault set $\calE_\mathrm{ref}$, we can always evaluate whether adding a CNOT gate to the circuit under construction preserves $t$-distinctness of the constructed fault set and the reference fault set. 
At any stage of the search, we can filter the CNOT gates that would alter the fault set of the circuit under construction, such that it would not be t-distinct from the reference circuit.
Among the remaining candidate CNOT gates, we can select the one that optimizes the cost function.
In practice, we first select a CNOT gate that locally optimizes the cost function and check whether it preserves t-distinctness. 
If it does not, the next best candidate is selected. 
This avoids many unnecessary and costly fault-tolerance checks up front.
This is illustrated in panel b of~\Cref{fig:algorithm_sketch}.

This search may get stuck in regions of the search space where no CNOT placement maintains $t$-distinctness. In other words, all the CNOTs we could place to prepare the desired logical state cause an overlap between the reference fault set and the current fault set of the circuit we are constructing at that moment. In this case, we have to backtrack by undoing previously chosen CNOT gates until the search can continue.
This situation is illustrated in panel c of~\Cref{fig:algorithm_sketch}.
The resulting algorithm is sketched in~\Cref{alg:heuristic}.

\begin{algorithm}[t]
\caption{Fault Set Guided Synthesis}\label{alg:heuristic}
\KwIn{Fault set $\mathcal{E}_{\text{ref}}$, fault-distance $t$, \\check matrix $H_X$, cost function $h$}
\KwOut{State preparation circuit with a $t$-distinct fault set to $\mathcal{E}_{\text{ref}}$}

$\mathcal{C} \gets$ empty circuits on $n$ qubits\;
$\calE \gets \{X_i \mid i \in [n]\} \cup \{Z_i \mid i \in [n]\}$\;
$\mathcal{B} \gets \emptyset$ \Comment*[r]{blocked CNOT gates}
$H\gets H_X$\;

\While{$H$ is not reduced}{
    $\mathrm{candidates} \gets \mathrm{argmin}_{\{\mathrm{CX}_{ij}\mid i,j\in[n]\}} h(\mathrm{CX}_{ij}, H)$\;
    
    \ForEach{$\mathrm{CX}_{ij} \in \mathrm{candidates} \setminus \mathcal{B}$}{
        $\calE' \gets \calE \cup \{\calC (X_iX_j) \calC^{-1}\} \cup \{\calC (Z_iZ_j) \calC^{-1}\}$\;
        \If{$\mathcal{E}'$ is $t$-distinct to $\mathcal{E}_{\text{ref}}$}{
            $\calC \gets \calC \circ \mathrm{CX}_{ij}$\;
            $H[:, j] \gets H[:, i] + H[:, j]$\;
            $\calE \gets \calE'$\;
            \textbf{Break}\;
        }
    }
    \If{no valid CNOT gates can be added}{ \Comment{backtrack}
        Remove last gate $cx_{\text{last}}$\;
        Add $cx_{\text{last}}$ to blocked list $\mathcal{B}$\;
    }
}
\Return{$\mathcal{C}$}
\end{algorithm}

Using this \emph{fault set guided synthesis} approach, we can construct the four circuits required for Steane-type FTSP:

\begin{enumerate}
   \item Construct $\calC_1$ using some state preparation circuit synthesis approach (e.g. Refs.~\cite{steaneFastFaulttolerantFiltering2004,zenQuantumCircuitDiscovery2025, pehamAutomatedSynthesisFaultTolerant2025, websterHeuristicOptimalSynthesis2025}).
   \item Construct $\calC_2$ using fault set guided synthesis with respect to $\calE_X(\calC_1)$.
   \item Construct $\calC_3$ using fault set guided synthesis with respect to $\calE_Z(\calC_1) \cup \calE_Z(\calC_2)$.
   \item Construct $\calC_4$ using fault set guided synthesis with respect to $\calE_X(\calC_3)$ and $\calE_Z(\calC_1) \cup \calE_Z(\calC_2)$.
\end{enumerate}

The construction of $\calC_4$ imposes the most constraints, and in practice, this leads to a significant increase in the number of backtracks.
In larger codes, the space of valid state preparation circuits expands significantly, making fault-set-guided synthesis more difficult. 
Furthermore, higher distances demand costlier t-distinctness checks as the fault sets to be checked can get large.

To help the search and avoid getting stuck in difficult parts of the search space, we employ a few strategies:

\begin{table*}[h!tbp]
    \centering
    \caption{Comparison of fault-tolerant state preparation circuit synthesis methods.}
    \label{tab:circuit_metrics}
    \footnotesize\setlength{\tabcolsep}{2.5pt}
    \begin{tabular}{l *{22}{c}}
    \toprule
      \multirow{3}{*}{Code} & \multicolumn{4}{c}{Ref.~\cite{pehamAutomatedSynthesisFaultTolerant2025}} & \multicolumn{4}{c}{Ref.~\cite{forlivesiFlagOriginModular2025}} & \multicolumn{12}{c}{Fault Set Guided Synthesis}                                                                                                                                                                                    \\
      \cmidrule(lr){2-5} \cmidrule(lr){6-9} \cmidrule(lr){10-21} 
                            & \multicolumn{2}{c}{Total}                                                & \multicolumn{2}{c}{Error}                                      & \multicolumn{2}{c}{Total}     & \multicolumn{2}{c}{Error}
                            & \multicolumn{2}{c}{$\calC_1$}                                            & \multicolumn{2}{c}{$\calC_2$}                                  & \multicolumn{2}{c}{$\calC_3$} & \multicolumn{2}{c}{$\calC_4$} & \multicolumn{2}{c}{Total} & \multicolumn{2}{c}{Error}                                                                                                              \\
    \cmidrule(lr){2-3} \cmidrule(lr){4-5} \cmidrule(lr){6-7} \cmidrule(lr){8-9} \cmidrule(lr){10-11}
    \cmidrule(lr){12-13} \cmidrule(lr){14-15} \cmidrule(lr){16-17} \cmidrule(lr){18-19} \cmidrule(lr){20-21}
                            & \#CX                                                                     & D                                                              & $p_l$                         & $r_A$                         & \#CX                      & D  & $p_l$              & $r_A$  & \#CX & D & \#CX & D & \#CX & D & \#CX & D & \#CX         & D           & $p_l$              & $r_A$ \\  \midrule
    $\llbracket17,1,5\rrbracket$            & \textbf{71}  & 22 & $1.4\times10^{-7}$  & 0.92          & 74           & 17 & $4.4\times10^{-7}$         & \textbf{0.92} & 23 & 5 & 23 & 4 & 23 & 5 & 23 & 4 & 92           & \textbf{7}  & $\mathbf{8.0\times10^{-8}}$  & 0.79          \\
    $\llbracket19,1,5\rrbracket$            & 120          & 25 & $1.0\times10^{-7}$  & \textbf{0.86} & -            & -  & -                          & -               & 27 & 4 & 27 & 4 & 27 & 6 & 27 & 5 & \textbf{108} & \textbf{8}  & $\mathbf{9.2\times10^{-8}}$  & 0.76          \\
    $\llbracket25,1,5\rrbracket$            & 127          & 31 & $7.4\times10^{-8}$  & 0.85          & \textbf{92}  & 20 & $3.5\times10^{-7}$          & \textbf{0.91} & 28 & 5 & 28 & 5 & 28 & 5 & 28 & 5 & 112          & \textbf{7}  & $\mathbf{5.8\times10^{-8}}$  & 0.72          \\
    $\llbracket20,2,6\rrbracket$            & -            & -  & -                   & -               & \textbf{145} & 51 & $1.6\times10^{-6}$          & \textbf{0.84} & 40 & 7 & 36 & 6 & 37 & 6 & 42 & 7 & 155          & \textbf{9}  & $\mathbf{8.6\times10^{-8}}$  & 0.72          \\
    $\llbracket31,1,7\rrbracket$            & 421          & 46 & $2.84\times10^{-9}$ & 0.56          & \textbf{211} & 55 & $4.1\times10^{-8}$          & \textbf{0.77} & 46 & 5 & 52 & 6 & 57 & 6 & 61 & 9 & 216          & \textbf{11} & $\mathbf{1.1\times10^{-9}}$  & 0.62          \\
    $\llbracket37,1,7\rrbracket$            & -            & -  & -                   & -               & -            & -  & -                          & -               & 60 & 5 & 60 & 6 & 77 & 7 & 77 & 9 & \textbf{274} & \textbf{11} & $\mathbf{1.0\times10^{-9}}$  & \textbf{0.55} \\
    \bottomrule
    \end{tabular}
    \vspace{1ex}
\begin{flushleft}
{\footnotesize \textbf{Legend:} \#CX: CNOT count; D: CNOT depth, i.e., the number of layers of parallel CNOT gates (for Ref.~\cite{forlivesiFlagOriginModular2025} computed from the published circuits); $p_l$, $r_A$: logical error rate and acceptance rate for physical error rate $p=10^{-3}$; '-': no circuit available. For each code, bold entries mark the best total CNOT count, depth, logical error rate, and acceptance rate among the compared methods.}
\end{flushleft}
\vspace*{-5mm}
\end{table*}

\begin{itemize}
    \item \emph{Avoid early conflicts by hard structural constraints.}
    
    Besides the fault set of the reference circuit, we know the exact structure of the circuit itself. If the last two layers of the reference circuit contain the substructure $CX_{ij}CX_{jk}$, then we know that $X_iX_jX_k$ will be in the fault set (assuming it is not reducible by a stabilizer). We can therefore enforce that the same structure cannot be used for the circuit under construction without performing any fault-tolerance checks.

    It may also be beneficial to avoid structures that do not directly cause conflicts but may lead to them later. 
    For example, it can be enforced that the last layer of CNOTs of the circuit under construction cannot have any CNOTs in common with the reference circuit. 
    The idea is that such constraints increase the likelihood that errors propagate similarly in both circuits in the last layer.
    
    \item \emph{Restart the search if the algorithm fails to make progress.}
    
    A typical technique in many solvers~\cite{biereHandbookSatisfiability2009} is to apply restarts, which help the search escape from unpromising parts of the search space where a choice far up the search tree causes the conflict. In our specific case, we can set a backtracking limit and restart the search once it is reached.

    \item \emph{Apply random perturbations to the search.}
    
    If multiple CNOTs have the same costs, one can be picked at random. 
    Additionally, a CNOT that has a (locally) higher cost might actually be a better choice for fault tolerance. We can increase the likelihood of selecting such CNOTs by randomly selecting lower-cost CNOTs with a probability parameterizing the search.

    \item \emph{Apply row operations to the input check matrix.}

    Since the check matrix represents a stabilizer group, row operations can be applied to it without altering the underlying group.
    The cost function depends on the specific shape of the check matrix, however, and by extension, so does the search.
    We can therefore try to guide the search to favor synthesizing different CNOT circuits by changing the matrix representation.
    In practice, a good heuristic is to bring the matrix into reduced row-echelon form and synthesize the circuit from that. 
\end{itemize}

Such heuristics help to remove large parts of the search space early on.

\section{Evaluation}
\label{sec:eval}

We implemented the proposed fault-set–guided synthesis of Steane-type FTSP circuits in the open-source error-correction library MQT~QECC~\cite{berentMQTQECC2025}, which is part of the Munich Quantum Toolkit~\cite{willeMQTHandbookSummary2024}. 
Using this implementation, we generated Steane-type FTSP circuits for various $d=5$, $d=6$, and $d=7$ CSS codes. 
In this section, we evaluate these circuits in terms of CNOT count, depth, logical error rate, and acceptance probability using Monte Carlo simulations.

We compare the proposed method with Ref.~\cite{pehamAutomatedSynthesisFaultTolerant2025}, which ensures fault tolerance by measuring a small set of stabilizers after non-fault-tolerant unitary state preparation, and with Ref.~\cite{forlivesiFlagOriginModular2025}, which augments the unitary preparation with flag gadgets that detect errors during initialization. In both cases, fault tolerance is achieved by post-selecting on all measurement outcomes being $+1$.

\begin{figure*}
    \centering
    \includegraphics[width=\linewidth]{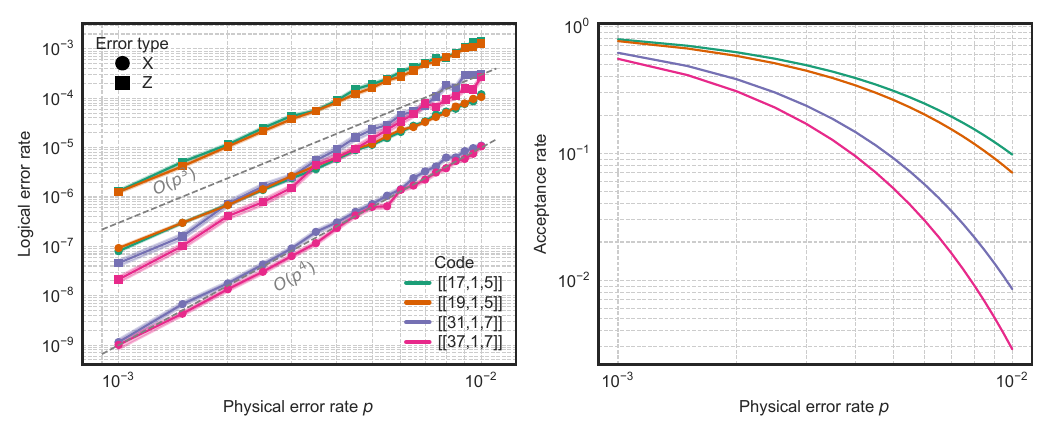}
    \caption{Logical error and acceptance rates for Steane-type fault-tolerant state preparation circuits for the logical $\ket{0}_L$ state constructed with our methods for various $d = 5$ and $d=7$ CSS codes under circuit-level noise using a LUT decoder.}
    \label{fig:css-state-sim}
\end{figure*}

\subsection{Circuits}
\label{sec:circuits}

\Cref{tab:circuit_metrics} shows CNOT gate count and depth for the Steane-type FTSP circuits for various CSS codes generated using different fault-tolerant state preparation methods.
The \enquote{Fault Set Guided Synthesis} columns refer to circuits synthesized using the construction described in~\Cref{sec:heuristic}.
The proposed method synthesizes four copies of a state so as to illustrate the variance in circuit size and depth, we list the circuits $\calC_1$, $\calC_2$, $\calC_3$, and $\calC_4$ required for fault-tolerant Steane-type state preparation.

For the distance-5 codes, \Cref{tab:circuit_metrics} shows that only minor adjustments to the reference circuits are required to achieve fault tolerance, with all designs exhibiting very similar CNOT counts and depths. This aligns with the analysis in \Cref{sec:17-1-5} and \Cref{sec:19-1-5}, which found that fault tolerance can be obtained through small modifications to the original circuit structures. For instance, the optimized $\llbracket17,1,5\rrbracket$ circuit improves both metrics, while the $\llbracket19,1,5\rrbracket$ circuit incurs only a slight increase in depth.

For distance-7 codes, the synthesis task becomes more challenging due to the larger circuit sizes and fault sets. In particular, ensuring $3$-distinctness for the $Z$-fault set of the fourth circuit is difficult, and a naive search often gets stuck. As described in~\Cref{sec:heuristic}, we mitigate this using additional heuristics; for the $d=7$ codes in \Cref{tab:circuit_metrics}, we synthesize $\mathcal{C}_4$ starting from the reduced row-echelon form of the check matrix to guide the search.

Overall, the circuits generated by the proposed method compare favorably to Ref.~\cite{pehamAutomatedSynthesisFaultTolerant2025}, as fault-set–guided synthesis yields more compact circuits in every instance except for the $\llbracket17,1,5\rrbracket$ code, where Ref.~\cite{pehamAutomatedSynthesisFaultTolerant2025} uses slightly fewer CNOT gates. In contrast, the “flag-at-origin’’ construction of Ref.~\cite{forlivesiFlagOriginModular2025} consistently requires fewer CNOT gates but incurs a significantly higher depth overhead.

In the flag-at-origin scheme, circuits must be synthesized from the reduced row-echelon form of the check matrix and further augmented with flag gadgets, both of which increase depth. In our construction, the overall depth is determined by the maximum depth among the four state-preparation circuits and by two additional layers of transversal CNOTs for verification. As a result, the Steane-type FTSP protocol based on fault-set–guided synthesis achieves comparatively low depth.

\subsection{Simulations}
\label{sec:simulation}

We simulated the synthesized circuits using the openly available stabilizer simulator Stim~\cite{gidneyStimFastStabilizer2021}.

We employ a depolarizing noise model parameterized by a physical noise strength $p$:
\begin{itemize}
    \item Two-qubit gates are followed by two-qubit depolarizing noise of strength $p$.
    \item Initializations and measurements flip with probability $\tfrac{2}{3}p$.
    \item Idle qubits in each circuit layer are subject to single-qubit depolarizing noise of strength $p/100$, and we assume gates can be executed in parallel with no crosstalk.
\end{itemize}

To estimate the logical $X$ error rate, we simulate the construction in~\Cref{fig:steane-ftsp-opt} for each code. 
Shots that indicate an error present on one of the three ancilla circuits are discarded.
In kept shots, we perform one round of perfect error correction on the remaining state  using a lookup-table decoder, measure the state in the standard basis, and check which eigenspace of the $Z$ operator it ended up in. 
A logical $-1$ measurement indicates an error.

Getting an estimate on the $Z$ error rate is trickier; a logical $Z$ error leaves $\ket{0}_L$ unchanged.
While $\ket{0}_L$ cannot directly be subject to logical $Z$ errors, high-weight $Z$ errors can still be an issue when using the prepared state as an ancilla in Steane-type error correction, where low-weight $Z$ faults during preparation can cause a high-weight correction on the data state.
Therefore, if preparation of the $\ket{0}_L$ is not strictly fault-tolerant, such interactions can compromise the performance of any FTQC protocol that uses the prepared states.

We use a trick to estimate the logical error rate scaling for $Z$ errors. 
To estimate $Z$ errors on the $\ket{0}_L$, we can use it as an ancilla for Steane-type error correction. 
We initialize another logical qubit in the $\ket{+}_L$ state and subject it to single-qubit depolarizing noise of strength $p$. 
Afterward, we copy the $Z$ error to the prepared $\ket{0}_L$ state with a transversal CNOT and measure out the state in the $X$ basis.
The syndrome is then decoded, and the correction applied to the $\ket{+}_L$ state.
We can then estimate the logical $Z$ error rate as before by measuring out the $\ket{+}_L$ state in the $X$ basis. 
If the $\ket{0}_L$ state is prepared strictly fault-tolerantly, then the correction applied to the $\ket{+}_L$ state due to the $Z$ syndrome should not change the logical error rate scaling as events of weight $t$ occur with probability $O(p^t)$, i.e., on the same order as in the single-qubit depolarizing channel.

The LUT decoder used in our simulations covers the entire syndrome space.
Lookup table entries are populated by enumerating errors in order of increasing weight, with lower-weight errors taking precedence.
For each syndrome, the first error encountered at its minimum weight is stored as the correction.
This means that for syndromes corresponding to uniquely correctable errors (weight $\leq \lfloor (d-1)/2\rfloor$), the LUT applies the unique minimum-weight correction.
For higher-weight syndromes, most notably weight-$(d/2)$ syndromes in the even-distance $\llbracket20,2,6\rrbracket$ code, where the code detects but cannot uniquely correct, the LUT applies one representative of the degenerate set of minimum-weight errors, with the specific choice determined by the iteration order.
No additional postselection is applied beyond the four-ancilla check of the FTSP protocol, so events in which the LUT applies an incorrect representative contribute to the reported logical error rate.

\Cref{fig:css-state-sim} shows the resulting logical error rates and acceptance rates of the simulated color code circuits for distances 5 and 7 generated by the fault set guided synthesis.
Generally, we can see that error rates scale as $O(p^{\ceil{d/2}})$ for the respective codes.
The higher probability of logical $Z$ errors is caused by the additional errors on the Steane-type error correction gadget used to estimate the logical $Z$ error rates. 

Besides the color code simulations, \Cref{tab:circuit_metrics} additionally summarizes the logical error rates and acceptance probabilities of the rotated surface code $\llbracket 25,1,5 \rrbracket$ and the $\llbracket 20,2,6\rrbracket$~code for the proposed state preparation methods, alongside the approaches of Ref.\cite{pehamAutomatedSynthesisFaultTolerant2025} and Ref.\cite{forlivesiFlagOriginModular2025}, evaluated at a physical error rate of $10^{-3}$. Compared to Ref.\cite{forlivesiFlagOriginModular2025}, the proposed method exhibits consistently lower acceptance rates in simulation. Relative to Ref.\cite{pehamAutomatedSynthesisFaultTolerant2025}, the acceptance rate is slightly reduced for distance-5 codes, while a clear improvement is observed for the distance-7 code. 
Although the proposed constructions exhibit very low circuit depths and a reduced number of CNOT gates compared to Ref.\cite{pehamAutomatedSynthesisFaultTolerant2025}, as well as only a marginally higher CNOT count than Ref.\cite{forlivesiFlagOriginModular2025}, this does not translate into higher acceptance rates. Instead, the lower acceptance rates are an expected consequence of the verification strategy employed. Since all ancilla qubits are fully measured and errors on the data can propagate throughout the gadget, acceptance effectively post-selects on the absence of errors across the entire circuit, up to acceptable cancellations. This stringent post-selection naturally reduces the acceptance rate but simultaneously suppresses logical errors. As a result, the lower acceptance rate directly correlates with improved logical error performance. Indeed, the proposed method achieves logical error rates that are roughly an order of magnitude lower than those reported in Ref.\cite{forlivesiFlagOriginModular2025} on average, with larger gaps for higher-distance codes, and performs similarly to Ref.\cite{pehamAutomatedSynthesisFaultTolerant2025} across all considered cases, with the exception of the $\llbracket20,2,6\rrbracket$ code. 
Overall, these results highlight a clear trade-off between acceptance rate and logical error suppression, under which the proposed constructions are preferable in settings where a lower acceptance rate is acceptable in exchange for significantly improved fault-tolerance.

The synthesis methodology operates directly on the check matrices $H_X$
and $H_Z$ of the code and does not exploit any specific structural
properties such as geometric locality, code symmetries, or the
transversality of particular gates. The choice of color codes in
\Cref{sec:states} reflects their suitability for illustrating the fault-set
arguments, and the predominance of topological codes in \Cref{tab:circuit_metrics}
reflects their relevance to current experimental platforms; the method
itself is code-agnostic. The same procedure applies to other CSS code
families relevant for fault-tolerant computation, including
triorthogonal codes used for magic state distillation, subject to the
same scaling considerations discussed in \Cref{sec:heuristic}.

\section{Conclusion}
\label{sec:conclusion}

In this work, we have investigated the problem of Steane-type FTSP for arbitrary CSS codes in which a logical basis state of a code is prepared using four copies of a state, copying errors using transversal CNOT gates, and post-selecting on the trivial measurement result.
We have shown how this can be achieved by altering the circuit structure of the CNOT circuit used for state preparation, either by permuting commuting CNOT gates or by using on-the-fly construction of preparation circuits utilizing a backtracking approach.
Furthermore, we have implemented the proposed methods and have provided an open-source toolkit for constructing and evaluating Steane-type FTSP circuits.

While we have shown the efficacy of the proposed fault-set-guided construction on selected $d=5$ and $d=7$ codes, the need for explicit construction of large fault sets for higher-distance codes poses an obstacle to the scalability of this approach. 
Future work could aim to alleviate or optimize this by trying a combination of automated approaches developed recently~\cite{forlivesiFlagOriginModular2025,pehamAutomatedSynthesisFaultTolerant2025}.
It might be possible to use the error detection of these constructions to reduce the size of the fault sets involved.

This work provides an important step in the compilation of fault-tolerant quantum programs. On the one hand, the states generated by this repeat-until-success protocol might be interesting for near-term demonstrations of fault-tolerant quantum computing with slow measurement and no real-time classical feedback. 
On the other hand, high-quality states prepared using the optimized circuits might be of interest for fault-tolerant quantum computing with Steane-type error correction.

\section*{Data Availability}

The data that support the findings of this article are openly available~\cite{weilandtDataSynthesisFaulttolerant2026}.
The implementation of the proposed methods is available as part of the open-source library MQT~QECC~\cite{berentMQTQECC2025}.

\section*{Acknowledgements}

The authors would like to thank Sascha Heußen for initial discussions on generalizing Steane-type state preparation of the Golay code, Ludwig Schmid for insightful discussions and comments, and Lucas Berent for helpful comments on the initial version of this manuscript.

The authors acknowledge funding from the European Research Council (ERC) under the European Union’s Horizon 2020 research and innovation program (grant agreement No.\ 101001318), from the European Union’s Horizon Europe research and innovation program under grant agreement No.\ 101114305 (``MILLENION-SGA1'' EU Project), and from the Munich Quantum Valley (MQV, K5 and K7), which is supported by the Bavarian state government with funds from the Hightech Agenda Bayern Plus. Furthermore, this work was supported by the BMFTR under grant number 13N17298 (SYNQ) and the Deutsche Forschungsgemeinschaft (DFG, German Research Foundation) under grant numbers 563402549 and 563436708.

\bibliography{zotero,new_refs}
\end{document}